


%
%


\def\famname{
 \textfont0=\textrm \scriptfont0=\scriptrm
 \scriptscriptfont0=\sscriptrm
 \textfont1=\textmi \scriptfont1=\scriptmi
 \scriptscriptfont1=\sscriptmi
 \textfont2=\textsy \scriptfont2=\scriptsy \scriptscriptfont2=\sscriptsy
 \textfont3=\textex \scriptfont3=\textex \scriptscriptfont3=\textex
 \textfont4=\textbf \scriptfont4=\scriptbf \scriptscriptfont4=\sscriptbf
 \skewchar\textmi='177 \skewchar\scriptmi='177
 \skewchar\sscriptmi='177
 \skewchar\textsy='60 \skewchar\scriptsy='60
 \skewchar\sscriptsy='60
 \def\rm{\fam0 \textrm} \def\bf{\fam4 \textbf}}
\def\sca#1{scaled\magstep#1} \def\scah{scaled\magstephalf} 
\def\twelvepoint{
 \font\textrm=cmr12 \font\scriptrm=cmr8 \font\sscriptrm=cmr6
 \font\textmi=cmmi12 \font\scriptmi=cmmi8 \font\sscriptmi=cmmi6 
 \font\textsy=cmsy10 \sca1 \font\scriptsy=cmsy8
 \font\sscriptsy=cmsy6
 \font\textex=cmex10 \sca1
 \font\textbf=cmbx12 \font\scriptbf=cmbx8 \font\sscriptbf=cmbx6
 \font\it=cmti12
 \font\sectfont=cmbx12 \sca1
 \font\refrm=cmr10 \scah \font\refit=cmti10 \scah
 \font\refbf=cmbx10 \scah
 \def\twelverm{\textrm} \def\twelveit{\it} \def\twelvebf{\textbf}
 \famname \textrm 
 \voffset=.04in \hoffset=.21in
 \normalbaselineskip=18pt plus 1pt \baselineskip=\normalbaselineskip
 \parindent=21pt
 \setbox\strutbox=\hbox{\vrule height10.5pt depth4pt width0pt}}


\catcode`@=11

{\catcode`\'=\active \def'{{}^\bgroup\prim@s}}

\def\screwcount{\alloc@0\count\countdef\insc@unt}   
\def\screwdimen{\alloc@1\dimen\dimendef\insc@unt} 
\def\screwbox{\alloc@4\box\chardef\insc@unt}

\catcode`@=12


\overfullrule=0pt			
\voffset=.04in \hoffset=.21in
\vsize=9in \hsize=6in
\parskip=\medskipamount	
\lineskip=0pt				
\normalbaselineskip=18pt plus 1pt \baselineskip=\normalbaselineskip
\abovedisplayskip=1.2em plus.3em minus.9em 
\belowdisplayskip=1.2em plus.3em minus.9em	
\abovedisplayshortskip=0em plus.3em	
\belowdisplayshortskip=.7em plus.3em minus.4em	
\parindent=21pt
\setbox\strutbox=\hbox{\vrule height10.5pt depth4pt width0pt}
\def\makefootline{\baselineskip=30pt \line{\the\footline}}
\footline={\ifnum\count0=1 \hfil \else\hss\twelverm\folio\hss \fi}
\pageno=1


\def\boxit#1{\leavevmode\thinspace\hbox{\vrule\vtop{\vbox{
	\hrule\kern1pt\hbox{\vphantom{\bf/}\thinspace{\bf#1}\thinspace}}
	\kern1pt\hrule}\vrule}\thinspace}
\def\Boxit#1{\noindent\vbox{\hrule\hbox{\vrule\kern3pt\vbox{
	\advance\hsize-7pt\vskip-\parskip\kern3pt\bf#1
	\hbox{\vrule height0pt depth\dp\strutbox width0pt}
	\kern3pt}\kern3pt\vrule}\hrule}}


\def\put(#1,#2)#3{\screwdimen\unit  \unit=1in
	\vbox to0pt{\kern-#2\unit\hbox{\kern#1\unit
	\vbox{#3}}\vss}\nointerlineskip}

%
%
%
%
%
%
%

\def\\{\hfil\break}

\def\center{\leftskip=0pt plus 1fill \rightskip=\leftskip \parindent=0pt
 \def\textindent##1{\par\hangindent21pt\footrm\noindent\hskip21pt
 \llap{##1\enspace}\ignorespaces}\par}
\def\unnarrower{\leftskip=0pt \rightskip=\leftskip}
\def\thetitle#1#2#3#4#5{
 \font\titlefont=cmbx12 \sca2 \font\footrm=cmr10 \font\footit=cmti10
  \twelverm
	{\hbox to\hsize{#4 \hfill ITP-SB-#3}}\par
	\vskip.8in minus.1in {\center\baselineskip=1.44\normalbaselineskip
 {\titlefont #1}\par}{\center\baselineskip=\normalbaselineskip
 \vskip.5in minus.2in #2
	\vskip1.4in minus1.2in {\twelvebf ABSTRACT}\par}
 \vskip.1in\par
 \narrower\par#5\par\unnarrower\vskip3.5in minus2.3in\eject}
\def\paper\par#1\par#2\par#3\par#4\par#5\par{\twelvepoint
	\thetitle{#1}{#2}{#3}{#4}{#5}} 
\def\author#1#2{#1 \vskip.1in {\twelveit #2}\vskip.1in}
\def\ITP{Institute for Theoretical Physics\\
	State University of New York, Stony Brook, NY 11794-3840}


\def\sect#1\par{\par\ifdim\lastskip<\medskipamount
	\bigskip\medskip\goodbreak\else\nobreak\fi
	\noindent{\sectfont{#1}}\par\nobreak\medskip} 
\def\itemize#1 {\item{[#1]}}	
\def\vol#1 {{\refbf#1} }		 

\def\ref#1{\setbox0=\hbox{M}$\vbox to\ht0{}^{#1}$}


\def\NP #1 {{\refit Nucl. Phys.} {\refbf B{#1}} }
\def\PL #1 {{\refit Phys. Lett.} {\refbf{#1}} }
\def\PR #1 {{\refit Phys. Rev. Lett.} {\refbf{#1}} }
\def\PRD #1 {{\refit Phys. Rev.} {\refbf D{#1}} }


\hyphenation{pre-print}
\hyphenation{quan-ti-za-tion}

%
%

\def\on#1#2{{\buildrel{\mkern2.5mu#1\mkern-2.5mu}\over{#2}}}
\def\dt#1{\on{\hbox{\bf .}}{#1}}                
\def\ddt#1{\on{\hbox{\bf .\kern-1pt.}}#1}    
\def\slap#1#2{\setbox0=\hbox{$#1{#2}$}
	#2\kern-\wd0{\hbox to\wd0{\hfil$#1{/}$\hfil}}}
\def\sla#1{\mathpalette\slap{#1}}                
\def\bop#1{\setbox0=\hbox{$#1M$}\mkern1.5mu
	\vbox{\hrule height0pt depth.04\ht0
	\hbox{\vrule width.04\ht0 height.9\ht0 \kern.9\ht0
	\vrule width.04\ht0}\hrule height.04\ht0}\mkern1.5mu}
\def\bo{{\mathpalette\bop{}}}                        
\def~{\widetilde} 
\mathcode`\*="702A                  
\def\in{\relax\ifmmode\mathchar"3232\else{\refit in\/}\fi} 
\def\f#1#2{{\textstyle{#1\over#2}}}	   
\def\half{{\textstyle{1\over{\raise.1ex\hbox{$\scriptstyle{2}$}}}}}

\catcode`\^^?=13				    
\catcode128=13 \def €{\"A}                 
\catcode129=13 \def {\AA}                 
\catcode130=13 \def '{\c}           	   
\catcode131=13 \def ƒ{\'E}                   
\catcode132=13 \def "{\~N}                   
\catcode133=13 \def …{\"O}                 
\catcode134=13 \def †{\"U}                  
\catcode135=13 \def ‡{\'a}                  
\catcode136=13 \def ˆ{\`a}                   
\catcode137=13 \def ‰{\^a}                 
\catcode138=13 \def Š{\"a}                 
\catcode139=13 \def ‹{\~a}                   
\catcode140=13 \def Œ{\alpha}            
\catcode141=13 \def {\chi}                
\catcode142=13 \def Ž{\'e}                   
\catcode143=13 \def {\`e}                    
\catcode144=13 \def {\^e}                  
\catcode145=13 \def '{\"e}                
\catcode146=13 \def '{\'\i}                 
\catcode147=13 \def "{\`\i}                  
\catcode148=13 \def "{\^\i}                
\catcode149=13 \def •{\"\i}                
\catcode150=13 \def –{\~n}                  
\catcode151=13 \def —{\'o}                 
\catcode152=13 \def ˜{\`o}                  
\catcode153=13 \def ™{\^o}                
\catcode154=13 \def š{\"o}                 
\catcode155=13 \def ›{\~o}                  
\catcode156=13 \def œ{\'u}                  
\catcode157=13 \def {\`u}                  
\catcode158=13 \def ž{\^u}                
\catcode159=13 \def Ÿ{\"u}                
\catcode160=13 \def  {\tau}               
\catcode161=13 \mathchardef ¡="2203     
\catcode162=13 \def ¢{\oplus}           
\catcode163=13 \def £{\relax\ifmmode\to\else\itemize\fi} 
\catcode164=13 \def ¤{\subset}	  
\catcode165=13 \def ¥{\infty}           
\catcode166=13 \def ¦{\mp}                
\catcode167=13 \def §{\sigma}           
\catcode168=13 \def ¨{\rho}               
\catcode169=13 \def ©{\gamma}         
\catcode170=13 \def ª{\leftrightarrow} 
\catcode171=13 \def «{\relax\ifmmode\acute\else\expandafter\'\fi}
\catcode172=13 \def ¬{\relax\ifmmode\expandafter\ddt\else\expandafter\"\fi}
\catcode173=13 \def ­{\equiv}            
\catcode174=13 \def ®{\approx}          
\catcode175=13 \def ¯{\Omega}          
\catcode176=13 \def °{\otimes}          
\catcode177=13 \def ±{\ne}                 
\catcode178=13 \def ²{\le}                   
\catcode179=13 \def ³{\ge}                  
\catcode180=13 \def ´{\upsilon}          
\catcode181=13 \def µ{\mu}                
\catcode182=13 \def ¶{\delta}             
\catcode183=13 \def ·{\epsilon}          
\catcode184=13 \def ¸{\Pi}                  
\catcode185=13 \def ¹{\pi}                  
\catcode186=13 \def º{\beta}               
\catcode187=13 \def »{\partial}           
\catcode188=13 \def ¼{\nobreak\ }       
\catcode189=13 \def ½{\zeta}               
\catcode190=13 \def ¾{\sim}                 
\catcode191=13 \def ¿{\omega}           
\catcode192=13 \def À{\dt}                     
\catcode193=13 \def Á{\gets}                
\catcode194=13 \def Â{\lambda}           
\catcode195=13 \def Ã{\nu}                   
\catcode196=13 \def Ä{\phi}                  
\catcode197=13 \def Å{\xi}                     
\catcode198=13 \def Æ{\psi}                  
\catcode199=13 \def Ç{\int}                    
\catcode200=13 \def È{\oint}                 
\catcode201=13 \def É{\relax\ifmmode\cdot\else\vol\fi}    
\catcode202=13 \def Ê{\relax\ifmmode\,\else\thinspace\fi}
\catcode203=13 \def Ë{\`A}                      
\catcode204=13 \def Ì{\~A}                      
\catcode205=13 \def Í{\~O}                      
\catcode206=13 \def Î{\Theta}              
\catcode207=13 \def Ï{\theta}               
\catcode208=13 \def Ð{\relax\ifmmode\bar\else\expandafter\=\fi}
\catcode209=13 \def Ñ{\overline}             
\catcode210=13 \def Ò{\langle}               
\catcode211=13 \def Ó{\relax\ifmmode\{\else\ital\fi}      
\catcode212=13 \def Ô{\rangle}               
\catcode213=13 \def Õ{\}}                        
\catcode214=13 \def Ö{\sla}                      
\catcode215=13 \def ×{\relax\ifmmode\check\else\expandafter\v\fi}
\catcode216=13 \def Ø{\"y}                     
\catcode217=13 \def Ù{\"Y}  		    
\catcode218=13 \def Ú{\Leftarrow}       
\catcode219=13 \def Û{\Leftrightarrow}       
\catcode220=13 \def Ü{\relax\ifmmode\Rightarrow\else\sect\fi}
\catcode221=13 \def Ý{\sum}                  
\catcode222=13 \def Þ{\prod}                 
\catcode223=13 \def ß{\widehat}              
\catcode224=13 \def à{\pm}                     
\catcode225=13 \def á{\nabla}                
\catcode226=13 \def â{\quad}                 
\catcode227=13 \def ã{\in}               	
\catcode228=13 \def ä{\star}      	      
\catcode229=13 \def å{\sqrt}                   
\catcode230=13 \def æ{\^E}			
\catcode231=13 \def ç{\Upsilon}              
\catcode232=13 \def è{\"E}    	   	 
\catcode233=13 \def é{\`E}               	  
\catcode234=13 \def ê{\Sigma}                
\catcode235=13 \def ë{\Delta}                 
\catcode236=13 \def ì{\Phi}                     
\catcode237=13 \def í{\`I}        		   
\catcode238=13 \def î{\iota}        	     
\catcode239=13 \def ï{\Psi}                     
\catcode240=13 \def ð{\times}                  
\catcode241=13 \def ñ{\Lambda}             
\catcode242=13 \def ò{\cdots}                
\catcode243=13 \def ó{\^U}			
\catcode244=13 \def ô{\`U}    	              
\catcode245=13 \def õ{\bo}                       
\catcode246=13 \def ö{\relax\ifmmode\hat\else\expandafter\^\fi}
\catcode247=13 \def÷{\relax\ifmmode\tilde\else\expandafter\~\fi}
\catcode248=13 \def ø{\ll}                         
\catcode249=13 \def ù{\gg}                       
\catcode250=13 \def ú{\eta}                      
\catcode251=13 \def û{\kappa}                  
\catcode252=13 \def ü{\half}     		 
\catcode253=13 \def ý{\Gamma} 		
\catcode254=13 \def þ{\Xi}   			
\catcode255=13 \def ÿ{\relax\ifmmode{}^{\dagger}{}\else\dag\fi}


\def\ital#1Õ{{\it#1\/}}	     
\def\un#1{\relax\ifmmode\underline#1\else $\underline{\hbox{#1}}$
	\relax\fi}

\def\tdt#1{\on{\hbox{\bf .\kern-1pt.\kern-1pt.}}#1}   
\def\({\eqno(}
\def\li{\eqalignno}
\def\refs{\sect{REFERENCES}\par\medskip \frenchspacing 
	\parskip=0pt \refrm \baselineskip=1.23em plus 1pt
	\def\ital##1Õ{{\refit##1\/}}}


\def\õ#1{
	\screwcount\num
	\num=1
	\screwdimen\downsy
	\downsy=-1.5ex
	\mkern-3.5mu
	õ
	\loop
	\ifnum\num<#1
	\llap{\raise\num\downsy\hbox{$õ$}}
	\advance\num by1
	\repeat}
\def\upõ#1#2{\screwcount\numup
	\numup=#1
	\advance\numup by-1
	\screwdimen\upsy
	\upsy=.75ex
	\mkern3.5mu
	\raise\numup\upsy\hbox{$#2$}}


\catcode`\|=\active \catcode`\<=\active \catcode`\>=\active 
\def|{\relax\ifmmode\delimiter"026A30C \else$\mathchar"026A$\fi}
\def<{\relax\ifmmode\mathchar"313C \else$\mathchar"313C$\fi}
\def>{\relax\ifmmode\mathchar"313E \else$\mathchar"313E$\fi}


\paper

COVARIANT FIELD THEORY\\FOR SELF-DUAL STRINGS

\author{N. Berkovits\footnote{${}^1$}{
	Internet address: nberkovi@ift.unesp.br.}}
	{Inst. de F«{\i}sica Te«orica, Univ.¼Estadual Paulista\\
	Rua Pamplona 145, S÷ao Paulo, SP 01405-900, Brasil}
\author{W. Siegel\footnote{${}^2$}{
	Internet address: siegel@insti.physics.sunysb.edu.}}\ITP

97-19, IFT-P.026/97

March 21, 1997

We give a gauge and manifestly SO(2,2) covariant formulation of the field
theory of the self-dual string.  The string fields are gauge connections
that turn the super-Virasoro generators into covariant derivatives.

Ü1. Introduction

There are three known kinds of string theories:  Those with critical
(uncompactified) dimension: (1) D=26 (``N=0"), which has various
fundamental problems (divergences, tachyons, no fermions, etc.); (2) D=10
(``N=1"), which is now thought to be a misleading formulation of a D=11
theory that includes supermembranes; and (3) D=4 [1] (``N=2" [2]), which
describes self-dual massless theories in 2 space and 2 time dimensions
[3,4].  (Note that critical/uncompactified dimension characterizes a string 
theory since, by embedding one string into another, the number of
worldsheet supersymmetries can be altered. It is not clear if the results
in this paper will be useful for N=2 strings which come from embeddings
of the D=26 and D=10 strings.)

The last type of string (the topic of this paper), because of its 2 time
dimensions, has lent itself to various interpretations.  Clearly unitarity 
cannot be applied in the usual way, and usually is ignored.  4D Lorentz
invariance is not manifest in the usual N=2 formulation, and therefore
also has largely been neglected, even though the self-dual field theories
with which it is identified have a Lorentz covariant definition.  Directly
related to the loss of manifest Lorentz invariance is the loss of gauge
invariance:  The N=2 formulations correspond to certain light-cone
gauges, which are not always the best choice for analyzing such theories,
particularly for such nonperturbative solutions as instantons.  Since spin
is ignored, statistics is also ignored; besides, unitarity and Lorentz
invariance are the usual justifications for their relation.  The N=2 string is
also equivalent [5] to the N=4 string [6], but although the latter
formulation is manifestly Lorentz covariant, its complicated ghost
structure has not been completely worked out, and therefore its existence
is seldom recognized.  Even dimensional analysis is a problem [7] since,
e.g., pure self-dual Yang-Mills contains no dimensionful coupling
constant, unlike the field theory action used in [3,4].

The precise definition of this string theory depends strongly on the
motivation for its consideration.  Up to now, all the work on the
noncovariant formulation of the self-dual string has been associated with
the fact that it implies classical field equations for self-dual Yang-Mills
theory or self-dual  gravity [3,4] (but not self-dual gravity coupled to
self-dual Yang-Mills [4])  in light-cone gauges.  These equations of motion
can be used to derive the classical equations of motion of wide classes of
integrable models in lower dimensions, as well as study certain
properties of solutions of the classical equations in four dimensions. 
However, most of the known 4D solutions (multi-instantons) require more
general gauges to be written explicitly [8].  (For example, explicit
n-Eguchi-Hanson and n-Taub-NUT solutions to self-dual gravity would
require solving 2n-th-order polynomial equations in this coordinate
system, and thus can be explicit only for n=1,2 [9].)  Furthermore, the
identification as the self-dual part of some non-self-dual theory at the
quantum level requires dimensional analysis to be consistent (e.g., for the
renormalization group).  In particular, the fact that these self-dual field
theories can be interpreted as (Wick rotations of) truncations of the
corresponding non-self-dual theories [10,11] implies that this string
theory actually can be used to help understand physical theories in 3+1
dimensions, and perform perturbative and nonperturbative calculations in
them.

For this purpose, it is useful to find a method of applying 2D conformal
field theory that preserves Lorentz and gauge invariances.  Traditionally,
the string field or wave function in any string theory has been assumed to
be a scalar (or at least a one-component field), with all excitations
described by its dependence on its arguments.   However, this is not a
physical requirement of the theory, but an assumption of the conformal
field theory description.  The same assumption is generally not made for
the quantum mechanics or quantum field theory of particles, and it is not
clear that such a requirement would aid in the evaluation of Feynman
diagrams.  Since the purpose of two-dimensional conformal field theory in
string theory is perturbation, one might consider calculational rules for
string S-matrices that allow for ``indices" in addition to the obvious
coordinates.   These indices are analogous to the Chan-Paton factors
which appear in open string theory and which have no conformal field
theory justification.  (Although it is true that Chan-Paton factors can be
associated with fermionic coordinates living on the worldsheet boundary,
their influence is always calculated by simply multiplying the
group-theory factors into the amplitude, rather than by calculating
worldline propagators for the fermions, etc.)   Just as the SO(32)
Chan-Paton factors of the light-cone-gauge open superstring are required
for SO(9,1) Lorentz  invariance,  the indices in the self-dual string are
needed for SO(2,2) Lorentz-invariance.  In earlier papers such indices
were associated with N=2 theories to restore Lorentz invariance (and
also allow supersymmetry) [10,7,11].  One immediate improvement over
the no-index formulation, even at the classical level and in the usual
light-cone gauges, is that the equations of motion can consistently
describe self-dual gravity coupled to self-dual Yang-Mills theory (see
section 2).

The purpose of this paper is to associate Lorentz indices with string fields
(or wave functions) in such a way as to give a string description of
self-dual theories while preserving gauge invariance and manifest
Lorentz invariance.  The usual string descriptions of these theories are
related to light-cone gauge choices (with the associated elimination of
auxiliary string fields).  The new formulation of the string theory, and its
relation to the conventional ones, is closely analogous to the known
treatment of the particle field theory describing just the massless fields. 
We therefore use, as our guide for covariantizing the string theory, the
covariant description of the particle theory, which we review in the
following section.  As for the particle field theory, two different
light-cone gauges are possible for the string field theory, corresponding
to the polynomial (cubic vertex) and nonpolynomial (Wess-Zumino-like)
formulations.  The string field theory has already been formulated in the
latter gauge [12], so we review it in section 3.  It is the formulation to
which we apply the covariantization, as described in section 4.  In the
final section we discuss supersymmetry and ghosts.

Ü2. Self-dual Yang-Mills theory

For purposes of perturbation theory, we can describe ordinary Yang-Mills
theory as a perturbation about self-dual Yang-Mills theory [13] (see [11]
for a light-cone approach):  We can write the Yang-Mills Lagrangian in
first-order form as [14]
$$ {\cal L} = G^{μ}F_{μ} +g^2 G^2 $$
 where $g$ is the usual Yang-Mills coupling, $G^{Œº}=G^{ºŒ}$ is an
anti-self-dual tensor, and
$$ F_{Œº} ­ [á_Œ{}^{À©},á_{ºÀ©}] = »_{(Œ}{}^{À©}A_{º)À©} +[A_Œ{}^{À©},A_{ºÀ©}] $$
 is the anti-self-dual-part of the usual Yang-Mills field strength $F$.  Here
$Œ=à$ is an SL(2,C) Weyl spinor index, and $ÀŒ=Àà$ is its complex conjugate,
in 3+1 dimensions.  To make the action real, we can Wick rotate to 2+2
dimensions, where $Œ$ becomes an SL(2) (SL(2,R)) index, and
$ÀŒ$ an SL(2)${}'$ index.  (SO(3,1)=SL(2,C), SO(2,2)=SL(2)$°$SL(2).)  Spinor
indices are raised and lowered with the SL(2) metric $C_{μ}$
(antisymmetric and Hermitian).  Note that although SL(2)=SU(1,1), the
SU(1,1) notation common in the N=2 string theory literature assumes a
particular representation where the U(1,1) metric is diagonal.  This differs
from the Majorana representation natural to SL(2) notation, where the
U(1,1) metric is antisymmetric.  In particular, in the diagonal
representation a spinor $^Œ$ satisfies $(^à)*=^¦$, while in the
antisymmetric (Majorana) representation $(^à)*=^à$.  In this paper
we'll generally refrain from using complex conjugation explicitly, since
this makes Wick rotation easier, so our notation can easily be specialized
to any representation.

Elimination of $G$ by its equation of motion produces the usual
Lagrangian, up to a total derivative.  (The action is then real in either 3+1
or 2+2.)  On the other hand, we can keep $G$, and treat $\hbar$
(${\cal L}£{\cal L}/\hbar$) and $g^2$ as independent expansion
parameters.  To lowest order in $g^2$ (i.e., $g=0$), we have a theory that
describes self-dual Yang-Mills theory, in the sense that $G$ is then a
Lagrange multiplier that enforces self-duality of $F$.  However, $G$ itself
is propagating, as required by Lorentz invariance:  Propagating helicity +1
in the self-dual part of $F$ requires propagating helicity $-$1 multiplying
it in the action.  This perturbation expansion in $g^2$ is natural in the
sense that the simplest tree and one-loop amplitudes in Yang-Mills
theory are those where (almost) all the external helicities are the same,
and the amplitudes become progressively more complicated as more
helicities change sign.  Similar remarks apply to self-dual gravity, where
the non-self-dual Lagrangian can be written in differential-form notation
as [14]
$$ {\cal L} = e^{ŒÀ©}\wedge e^º{}_{À©}\wedge 
	(d¿_{Œº} +û^2 ¿_{Œ}{}^{¶}\wedge ¿_{¶º}) $$
 where $e^{ŒÀº}=dx^m e_m{}^{ŒÀº}$ is the vierbein form (the analog of
$A$ above) and $¿^{Œº}=dx^m ¿_m{}^{Œº}$ is the anti-self-dual part of
the Lorentz connection form (the analog of $G$ above).  

In fact, almost all the amplitudes at $g=0$ or $û=0$ vanish, so this term in
the action is very similar to a kinetic term:  In the self-dual theories
described by the above actions, (1) all the tree amplitudes vanish on shell
except for the three-point (but it also vanishes on-shell in 3+1 dimensions
for kinematic reasons), and (2) all amplitudes vanish at more than one
loop, since $\hbar$ counting arguments show that any $L$-loop diagram
must have 1$-L$ external Lagrange multiplier lines.  (The classical action,
which goes as 1/$\hbar$, is linear in Lagrange multipliers, so in the
effective action the order in Lagrange multipliers is minus the order in
$\hbar$.)  Furthermore, in any of the supersymmetric versions of the
self-dual theories (N=1 supersymmetry or greater), all the one-loop
amplitudes also vanish.

From now on we restrict ourselves to the self-dual theories ($g=û=0$, or
lowest order in that perturbation expansion).  There are two light-cone
gauges for analyzing the self-duality condition $F_{μ}=0$ in Yang-Mills
theory (see [11] for an analysis at the quantum level): (1) a gauge
proposed by Yang [15], which gives field equations resembling a 2D
Wess-Zumino model, and (2) a gauge that gives a quadratic field equation,
found by Leznov, Mukhtarov, and Parkes [16].  In both cases, we first
choose the light-cone gauge and then solve the $F_{++}=0$ part of the
self-duality condition:
$$ \left. \matrix{ gauge¼A_{+À+} = 0 \hfill\cr 
	F_{++} = 0âÜâA_{+À-} = 0 \cr } \rightÕâÜâA_{+ÀŒ} = 0 $$
 The two cases differ in which of the remaining two equations is solved
as a constraint, and which is left as a field equation:  In the Yang case,
$$ \li{ F_{--} & = 0âÜâA_{-ÀŒ} = e^{-Ä}»_{-ÀŒ}e^Ä \cr
	F_{+-} & = 0âÜâ»_+{}^{ÀŒ}(e^{-Ä}»_{-ÀŒ}e^Ä) = 0 \cr } $$
 while in the LMP case
$$ \li{ F_{+-} & = 0âÜâA_{-ÀŒ} = »_{+ÀŒ}Ä \cr
	F_{--} & = 0âÜâ-iõÄ +(»_+{}^{ÀŒ}Ä)(»_{+ÀŒ}Ä) = 0 \cr } $$
 Note that in the LMP case the light-cone ``time" derivatives $»_{-ÀŒ}$
appear only in the kinetic term $õÄ$, while the Yang case is more like a
Wess-Zumino model, with such derivatives included in the interaction
term.

If we denote the surviving component of $G^{μ}$ by $ր$ ($G^{+-}$ in the
Yang case, $G^{--}$ in the LMP case), then the Lagrangian becomes just
$÷Ä$ times the $Ä$ field equation.  Note that $G$ (and thus $÷Ä$) has
engineering dimension 2, while $Ä$ is dimensionless.  Also, the Lorentz
transformations of $Ä$ and $÷Ä$ differ.  (This is especially clear in the LMP
case, where they even have different weights under the unbroken GL(1)
subgroup of the SL(2) acting on the undotted spinor indices.)  Thus, it is
not possible to write an action in terms of just $Ä$ that reproduces the
above field equations, without violating Lorentz invariance and
introducing a dimensionful coupling.  (Even with $ր$, Lorentz invariance is
not manifest:  The Lorentz transformations are nonlinear in the fields.) 
Furthermore, using a single field $Ä$ destroys the correspondence with
ordinary Yang-Mills theory, as described by a perturbation about the
self-dual theory.  Perturbatively, the (off-shell) tree graphs
agree, since they are the classical field equations.  The only difference is
the labeling of the external lines:  Calling one external line $ր$ and the
rest $Ä$ gives the same Feynman diagram as labeling all lines $Ä$
(although the interpretation is different).  However, the 1-loop graphs of
the single-field theory differ by a factor of 1/2, and it has
nonvanishing higher-loop graphs that have no apparent relation to
Yang-Mills theory.  As for open superstrings with different gauge groups,
the only difference in the theories is the index structure (in this case,
1-valued index vs.¼2-valued), but this makes all the difference in the
quantum corrections.

This analysis has been extended to self-dual gravity, and to self-dual
gravity coupled to Yang-Mills theory (as well as their supersymmetric
versions) [7].  In the case of gravity, the analog of the Yang gauge is the
Pleba«nski gauge [17], which in this case gives quadratic field equations. 
While the analog of the LMP gauge [7] again gives quadratic field
equations, they differ from the Pleba«nski gauge by the absence of ``$-$"
derivatives in the interaction term.  (In N=2 string theory, the Pleba«nski
gauge arises if world-sheet instantons are ignored, while the LMP-like
gauge follows if they are included [18].)  In both cases, the gravitational
3-point vertex is the square of the corresponding Yang-Mills one (4
derivatives instead of 2).  

In general, then, the differences between the various actions are rather
small, at least at the level of the propagator and 3-point vertices. 
(Higher-point vertices exist only for Yang-Mills theory, and only in the
Yang gauge, whether with or without a Lagrange multiplier.)  Explicitly,
the kinetic term in the lagrangian is always of the form
$$ {\cal L}_2 ¾ Ä_1 õ Ä_2 $$
 while the cubic term representing Yang-Mills coupling is
$$ {\cal L}_3 ¾ Ä_1 (»Ä_2)(»Ä_3) $$
 and that for gravitational coupling is
$$ {\cal L}_3' ¾ Ä_1 (»»Ä_2)(»»Ä_3) $$
 The explicit indices on the derivatives differ for Yang/Plebanski gauges
vs.¼LMP gauges; here we will instead focus on the difference with or
without Lagrange multiplers.  Without Lagrange multipliers:  (1) The
kinetic term has $Ä_1$ and $Ä_2$ the same in any such term, independent
of helicity (whether graviton, gluon, or their superpartners); (2) the
Yang-Mills coupling ${\cal L}_3$ also has all fields the same, namely 3
gluons (or a supersymmetric generalization); and (3) the gravitational
coupling ${\cal L}_3'$ has either 3 gravitons, or 2 gluons and 1 graviton (or
supersymmetric generalization).   On the other hand, when Lagrange
multipliers are introduced, each of the 3 kinds of terms is linear in them
(and thus either linear or quadratic in the usual fields).  In that case, it's
simpler to describe the couplings in terms of helicity: +2 for (self-dual)
graviton, $-$2 for its Lagrange multiplier, +1 for photon, etc.  Then
${\cal L}_2$ has fields with helicity summing to 0, ${\cal L}_3$ has any
fields with helicity summing to +1, and ${\cal L}_3'$ has any summing to
+2.  (This applies also to the supersymmetric cases.)

There are several levels of Lorentz invariance a description of a theory
can have:  (1) The highest is when the theory is described in terms of an
action that is manifestly Lorentz invariant.  (2) The next level, as results
for example when a noncovariant gauge is chosen or some auxiliary fields
are noncovariantly eliminated, is when there is an action that is still
invariant, but for which the Lorentz transformations are nonlinear (and
perhaps even nonlocal).  This is the case for the light-cone gauge actions
described above when the Lagrange multiplier fields are included.  (3) An
even lower level is that for the corresponding case when the Lagrange
multiplers are absent; the action is then not Lorentz invariant in any
sense, but the field equations are Lorentz covariant in the sense of the
previous level.  (4) The lowest level lacks any kind of Lorentz invariance
for even the field equations.  This is the case for the coupling of the
closed and open N=2 strings, as found in [4], which we now discuss in
more detail.  There the term ``self-dual" is loosely applied, since the
Pleba«nski equation gets a source term from the gluons, and thus no
longer describes self-dual gravity.  The resulting field equation has no
Lorentz covariant analog.  The vertices are exactly those described in the
previous paragraph.  On the other hand, the Lorentz covariant action with
Lagrange multiplers that we have discussed reproduces these vertices,
except for the different index structure.  The necessity for the Lagrange
multipliers for a covariant interpretation is clear from the covariant
actions given above:  For self-dual gravity coupled to self-dual
Yang-Mills, the actions given above (gravitationally covariantized for the
Yang-Mills terms) give the field equations
$$ S = eed¿ +GFâÜâ0 = \leftÓ \matrix{ ¶S/¶¿ & = & dee \cr
	¶S/¶G & = & F \cr
	¶S/¶A & = & áG \cr
	¶S/¶e & = & ed¿ +GF \cr } \right. $$
 where all indices are implict.  Thus, the self-duality equations for the
vierbein and Yang-Mills are unaffected (except for covariantization of
the latter), while the self-dual Yang-Mills energy-momentum tensor
$G_{Œº}F_{ÀŒÀº}$ appears in the field equation for $¿$.  This clearly
corresponds to the index structure described for the light-cone
Lagrangian terms described in the previous paragraph, where the fields
$e$, $A$, $G$, $¿$ have helicities +2, +1, $-$1, $-$2.  Thus, a simple
relabeling of fields has strong implications even at the level of classical
field equations.

Ü3. Noncovariant version of self-dual string field theory

In a paper by one of the authors [19],  it was shown how to construct an
open string field theory action for any critical N=2 superconformal 
representation. This action differs from the standard open string field
theory action [20], $ìQì +Âì^3$, in that it is built directly out of N=2
matter fields and does not require worldsheet ghosts. This is possible
since, after twisting, N=2 ghosts carry no central charge and decouple
from scattering amplitudes. This ghost-free description of N=2 strings
was developed by one of the authors with C. Vafa [12] and is extremely
useful for calculating N=2 scattering amplitudes [12,21].

In the ghost-free description of N=2 strings, it is useful to note  that any
critical N=2 representation contains generators of a ``small'' N=4
superconformal algebra.  For the self-dual representation of the N=2
string, the left-moving N=4 generators are:
$$ T = (»_z X^{ŒÀº})(»_z X_{ŒÀº}) +Æ^{ÀŒ¬º}»_z Æ_{ÀŒ¬º} $$
$$ G^{Œ¬º} = Æ^{À©¬º}»_z X^Œ{}_{À©} $$
$$ J^{¬Œ¬º} = Æ^{À©¬Œ}Æ_{À©}{}^{¬º} $$
 where $X^{ŒÀº}$ and $Æ^{ÀŒ¬º}$ are the usual string coordinates.  $Œ$
(SL(2)) and $ÀŒ$ (SL(2)${}'$) are the usual 4D Weyl spinor indices, while $¬Œ$
(the SL(2)${}''$ of $J^{¬Œ¬º}$) is the world-sheet internal index.  The $c=6$
N=2 generators are $T$, $G^{+¬-}$, $G^{-¬+}$,$ J^{¬+¬-}$ . After twisting
$T£T'=T+üJ^{¬+¬-}$, $T$, $G^{Œ¬-}$ and $J^{¬-¬-}$ carry spin two, $G^{Œ¬+}$ and
$J^{¬+¬-}$ carry spin one, and $J^{¬+¬+}$ carries spin zero.  Furthermore, 
$Æ^{ÀŒ¬+}$ is now spin zero while $Æ^{ÀŒ¬-}$ is spin one.  

In this formulation, only SL(2)${}'$ is completely preserved manifestly. 
Although in this paper we work in an N=2 formulation, we'll find that both
spacetime SL(2)'s can be preserved in the string field theory after adding
indices on the string field.  However, SL(2)${}''$ remains broken to the
usual local U(1) (or GL(1)) symmetry of the worldsheet, generated by
$J^{¬+¬-}$.   (The $¬Œ=¬à$ indices refer to the U(1) charge.)

As was described in [12], these generators can be used to compute
$N$-point scattering amplitudes on surfaces of genus (field-theory-loops)
$L$ and instanton number $n_I$ where $|n_I|² 2L-2+N$. The most relevant
scattering amplitude for open string field theory is the three-point tree
amplitude at zero instanton number, which is given by:
$$ Òì(z_1) (Q^+ ì(z_2))(Q^- ì(z_3))Ô \(3.1) $$
 where  $Q^Œ ì$ signifies the contour integral of spin-one $G^{Œ¬+}$
around the vertex operator $ì$ and $Ò¼Ô$ signifies the two-dimensional
correlation function on a sphere.  Note that this correlation function
vanishes unless the two zero-modes of $Æ^{ÀŒ¬+}$ are present. 

To be a physical vertex operator, $ì(X,Æ)$  must be U(1)-neutral (in the
N=2 topological  method, worldsheet U(1) charge is equal to spacetime
ghost number) and satisfy the linearized equation of motion
$Q^Œ Q_Œ ì=0$.  (This implies that $ì$ is a weight-zero N=2 primary
field.)  Unlike the usual vertex operator in string theory, $ì$ is defined
to be bosonic.  Note that the contour integral of $Q^+$ anticommutes with
the contour integral of $Q^-$, so (3.1) is invariant under the linearized
gauge transformation  $¶ì= Q^Œ ñ_Œ$. As in all open string theories, $ì$
carries Chan-Paton factors which will be suppressed throughout this
paper. 

Up to gauge transformations, the only momentum-dependent U(1)-neutral
vertex operator satisfying $Q^Œ Q_Œ ì=0$ is 
$ì=\exp(i k^{ŒÀŒ} X_{ŒÀŒ})$ where $k^{ŒÀŒ} k_{ŒÀŒ}=0.$ After performing
the correlation functions over the N=2 matter fields (remembering that
$Æ^{ÀŒ¬+}$ has a zero-mode), one finds that (3.1) produces the usual
three-point tree amplitude $k_2{}^{+ÀŒ}k_3{}^-{}_{ÀŒ}f^{I_1 I_2 I_3}$
where  $f^{I_1 I_2 I_3}$ is the structure constant for the Chan-Paton
factors and $k_r$ is the momentum of the $r$-th state.

To construct a string field theory action, it is natural to generalize the
on-shell vertex operator to an off-shell string field $ì$ which is an
arbitrary function of  $X(§)$, $Æ(§)$.  Note that the U(1) (GL(1)) charge of
$ì$ is related to the ghost number of the spacetime field.  If one chooses
the Majorana representation for SL(2)${}''$ spinors (implying that our
choice of $J^{¬+¬-}$ for the N=2 super-Virasoro algebra corresponds to a
GL(1) subgroup of this SL(2), rather than a U(1) subgroup), the reality
condition on the string field is the usual one:
$$ ì(§)ÿ = ì(¹-§) \(3.2) $$
 Note that in this representation,  twisting $T£T'=T+üJ^{¬+¬-}$ commutes
with hermitian conjugation. One can also ``Wick-rotate" this choice to the
diagonal representation for (the U(1,1) metric of) the SL(2)${}''$ spinors
(implying that $J^{¬+¬-}$ corresponds to a U(1) subgroup), but then
hermitian  conjugation must be accompanied by an SL(2) transformation 
to restore the original twist [19].

For the string field theory action to be correct, the quadratic term in the 
action should enforce the linearized equation of motion $Q^Œ Q_Œ ì=0$,
while the cubic term should produce the correct on-shell three-point
amplitude. Finally, the action should contain a gauge invariance whose
linearized form is $¶ì=Q^Œ ñ_Œ$.

The quadratic and cubic terms in the action are easily found to be of the
form
$$ üÇ\left[üi(Q^Œ ì)(Q_Œ ì) -\f13 ìÓQ^+ì,Q^-ìÕ\right] $$
 where ``$Ç$" means integration over all the modes of $X$ and $Æ$. 
However, there is no nonlinear version of the gauge transformation
$¶ì=Q^Œ ñ_Œ$ which leaves this action invariant. Knowing the equations
of motion for self-dual Yang-Mills in Yang gauge, an obvious guess for the
nonlinear generalization of $Q^Œ Q_Œ ì=0$ is $Q^-(e^{-ì}Q^+ e^{ì})=0$,
where multiplication of string fields is always performed using Witten's
half-string overlap. If $Ä$ is the component of $ì$ which depends only on
the zero-mode of $X$, then  $Q^-(e^{-ì}Q^+ e^{ì})$ contains the term
$Æ^{ÀŒ¬+} »^-{}_{ÀŒ} (e^{-Ä}Æ^{Àº¬+} »^+{}_{Àº} e^{Ä})$=
$Æ^{ÀŒ¬+}Æ_{ÀŒ}{}^{¬+} »^{-Àº} (e^{-Ä}»^+{}_{Àº} e^{Ä})$, which is
$Æ^{ÀŒ¬+}Æ_{ÀŒ}{}^{¬+}$ times the self-dual equation of motion in Yang gauge.

The action which produces this equation of motion is a straightforward
generalization of the Wess-Zumino model [22] where the two-dimensional
derivatives $»_z$ and $л_{Ðz}$ are replaced by $Q^+$ and $Q^-$. The string
field theory action is 
$$üÇ\left[ (e^{-ì} Q^+ e^{ì})(e^{-ì}Q^- e^{ì})
	-Ç_0^1 dt (e^{-tì}»_t e^{tì})Ó e^{-tì}Q^+ e^{tì},
	e^{-tì}Q^- e^{tì}Õ \right]  \(3.3) $$
 In addition to producing the correct linearized equations of motion and
three-point tree amplitude, this action contains the nonlinear gauge
invariance,
$$ ¶ e^{ì}= (Q^+ñ_+) e^ì  + e^ì (Q^-ñ_-) \(3.4) $$
 which generalizes the linearized gauge invariance $¶ì=Q^Œ ñ_Œ$.

Ü4. Lorentz covariance

In this section, we show how to ``covariantize'' the field theory action
for the self-dual representation of the N=2 string.  It is still unclear if the
covariantization procedure will be useful for other N=2 superconformal
representations.  The results of the previous section for the N=2 open
string are clearly analogous to those for self-dual Yang-Mills in the Yang
gauge, with the identification
$$ Q_Œ ª »_{ŒÀº} $$
 I.e., in the string field theory the dotted spinor indices are dropped, the
antisymmetry of the SL(2)${}'$ metric on those indices being replaced with
the anticommutativity of the ``BRST operators" $Q_Œ$.  (The dotted spinor
indices reappear if one expands out the $Æ^{ÀŒ¬+}$ dependence.) 

This suggests that to recover 4D Lorentz invariance, one needs to place a
two-valued index on the string field: 
$$ì_1 = ÷ì,ââì_2 = ì $$
 where $÷ì$ plays the role of the Lagrange multiplier and $ì$ plays the role
of the self-dual field. Furthermore, the string field action in Yang gauge
needs to be modified to $Ç÷ì Q^-(e^{-ì} Q^+ e^{ì})$.  Note that except for
the index structure and numerical (permutation) factors, this action has
the same quadratic and cubic terms as (3.3), and the same linearized
gauge transformation (see below).  The change in the index structure has
the effect of multiplyling the usual conformal field theory calculation by a
factor $¶_{1-n,L} 2^L$ where $n$ is the number of tilded vertex operators
and $L$ is the number of loops.

We can extend the analogy to the manifestly Lorentz covariant
formulation by proposing the new string field theory action
$$ S = ÇG^{Œº}F_{Œº},ââF_{Œº} = Óá_Œ,á_ºÕ,ââá_Œ = Q_Œ +A_Œ $$
 which is invariant under the gauge transformations
$$ á'_Œ = e^K á_Œ e^{-K},ââ
	G_{Œº}' = e^K G_{Œº}e^{-K} +á^© ¯_{©Œº} $$
 where $K$ is arbitrary and $¯_{©Œº}$ is symmetric in its indices.  $A_Œ$
and $G^{μ}$ are our new string fields, now carrying SL(2) spinor indices
as well as implicit Yang-Mills gauge-group indices (and with the same
string coordinates as arguments).  Note that the on-shell (surviving) field
strength $F_{ÀŒÀº}$ has no simple string field analog.  This is standard in
string field theory:  E.g., in N=0 open string field theory, the field strength
that vanishes on-shell ($Qì+ì¡ì$) is obvious, while the nonvanishing
string field strength has no local expression.

The string field theory action of the previous section can now be
rederived from this covariant action by the same methods described in 
section 2.  However, since the BRST operators $Q_Œ$ have nontrivial
kernels (in contrast to the partial derivatives $»_{ŒÀº}$), new gauge
invariances arise upon solving the constraints, in close analogy to 4D N=1
super Yang-Mills theory [19].  (See [23] for a review.)  In this case, the
gauge invariances arise because of the unphysical ``massive" fields,
absent in the discussion of section 2.

Explicitly, we find
$$ F_{++} = 0âÜâA_+ = 0 $$
in an appropriate gauge.  (I.e., $F_{++}=0$ implies $A_+$ is pure gauge.) 
However, this does not completely fix the gauge:  We are left with the
residual gauge invariance
$$ 0 = ¶A_+ = -Q_+ KâÚâK = Q_+ þ $$
 In particular, this applies for the gauge transformation of $G$.  

For the next step, in the Yang gauge,
$$ F_{--} = 0âÜâA_- = e^{-ì}Q_- e^ì $$
 This introduces the gauge invariance
$$ (e^ì)' = e^ñ e^ì,ââQ_- ñ = 0âÚâñ = Q_- Î $$
 The complete gauge transformations for $ì$ are now
$$ (e^ì)' = e^ñ e^ì e^{-K};ââK = Q_+ þ,âñ = Q_- Î $$
 Finally, the field equation is
$$ 0 = F_{+-} =  Q_+ (e^{-ì}Q_- e^ì) $$
 The Lagrangian then reduces to $֓F_{+-}$, where $֓=G_{+-}$.  Applying
these results, the gauge transformation for $֓$ is then
$$ ¶÷ì = [Q_+ þ,÷ì] +Q_+ ¯_{-+-} +Q_- ¯_{++-} +Óe^{-ì}Q_- e^ì,¯_{++-}Õ $$
 So, as claimed above, the linearized gauge transformations of  $Ç÷ìF_{+-}$
are of the form $¶ì=Q^Œ ñ_Œ$ and $¶÷ì=Q^Œ ÷ñ_Œ$.

On the other hand, we can also find a new light-cone string action by
going to an LMP gauge:
$$ F_{+-} = 0âÜâA_- = Q_+ ì $$
 This introduces the ÓAbelianÕ gauge invariance
$$ ¶ì = ñ,âñ = Q_+ Î $$
 The complete gauge transformation for $ì$ is now
$$ ¶ì = (Q_-þ +[Q_+þ,ì]) +Q_+ Î $$
 The field equation is now polynomial:
$$ 0 = F_{--} = -iQ^Œ Q_Œ ì +2(Q_+ ì)^2 $$
 and the Lagrangian is $֓F_{--}$, $֓=G_{++}$.  The gauge transformation
of $֓$ is
$$ ¶÷ì = [Q_+ þ,÷ì] +Q_+ ¯_{-++} +Q_- ¯_{+++} +ÓQ_+ ì,¯_{+++}Õ $$

Ü5. Supersymmetry and ghosts

The supersymmetric generalizations of self-dual theories have also been
analyzed [10,7].  Self-dual Yang-Mills theory can be treated as a
truncation of self-dual N=4 super Yang-Mills theory, where $A$ and $G$
are in the same supermultiplet.  (So can self-dual super Yang-Mills
theories for N<4.)  The component action in 2+2 dimensions (where all
fields are real) is 
$$ {\cal L} = üG^{Œº}F_{Œº} +Å^{iŒ}á_Œ{}^{Àº}_{iÀº}
	+·^{ijkl}(\f18 Ä_{ij}õÄ_{kl} +\f14 Ä_{ij}_k{}^{ÀŒ}_{lÀŒ}) $$
 where $i,j,k,l=1,...,4$ are the internal SL(4) indices of N=4 supersymmetry
(not to be confused with the Yang-Mills group indices, which are still
implicit), and $Ä_{ij}$ is antisymmetric.  Furthermore, self-dual Yang-Mills
theory, self-dual gravity, and self-dual gravity coupled to self-dual
Yang-Mills theory all can be treated as truncations of gauged self-dual
N=8 supergravity (with Yang-Mills gauge group SO(8)).  In light-cone
gauges, the vertices (and, of course, the propagators) are identical to the
nonsupersymmetric cases:  Spin appears effectively as an internal
symmetry index.

This helicity-independence of the couplings also has an explanation in
terms of N=2 strings:  Spectral flow is usually interpreted as allowing the
identification of states with different boundary conditions (which would
normally be associated with different spins, and thus different
statistics).  However, these states can be distinguished if they are
assigned different helicities.  (The assignment of helicities is somewhat
arbitrary in N=2 string theory; in fact, the usual continuous helicity
representations of the Poincar«e group [24] can be associated with these
self-dual theories, but only if one abandons the possibility of manifest
Lorentz covariance.)  So, rather than using spectral flow to say there's
only one state, it can be interpreted as a stronger version of
supersymmetry that  implies helicity-independence of the couplings.

As for the nonsupersymmetric case, the above component action can be
straightforwardly generalized to string field theory by dropping dotted
spinor indices and replacing $ȣQ$:
$$ {\cal L} = üG^{Œº}F_{Œº} +Å^{iŒ}á_Œ _{i}
	+·^{ijkl}(\f18 Ä_{ij}õÄ_{kl} +\f14 Ä_{ij}_k _l) $$
 where now $õ­üá^Œ á_Œ$.  The component fields describing helicities
+1, +1/2, 0, $-$1/2, $-$1 are now $A_Œ$, $_i$, $Ä_{ij}$, $Å^{iŒ}$,
$G^{μ}$.  Half of the supersymmetry transformations (those not
involving $F_{ÀŒÀº}$ explicitly) can also be generalized:
$$ \li{ ¶A_Œ & = ·^i{}_Œ _i \cr
	¶_i & = ·^{jŒ}á_Œ Ä_{ji} \cr
	¶Ä_{ij} & = -·_{ijkl}·^{kŒ}Å^l{}_Œ \cr
	¶Å^i{}_Œ & = ü·^{iº}G_{Œº} +ü·^m{}_Œ ·^{ijkl}[Ä_{jk},Ä_{lm}] \cr
	¶G_{Œº} & = -·^i{}_{(Œ}[Å^j{}_{º)},Ä_{ij}] \cr } $$

There is an interesting analogy between supersymmetry and the
Zinn-Justin-Batalin-Vilkovisky formalism:  The minimal field theory
Lagrangian for the nonsupersymmetric self-dual string, including
antifields and the ghosts for just the Yang-Mills gauge symmetry, is
$$ {\cal L} = G^{Œº}F_{Œº} +A*^Œ á_Œ C +C*C^2 $$
 This is very similar to the supersymmetric action (without antifields),
with the identification
$$ _i ª C,âÄ_{ij} ª C*,âÅ^{iŒ} ª A*^Œ $$
 This suggests a generalization of the GL(N) internal symmetry of
N-extended supersymmetry to SL(N|1).  (For N<4, the Lagrange multipliers
form a separate supermultiplet from the other fields, and the $·_{ijkl}$'s
can be absorbed.)  Because of the nontrivial cohomology of $Q_Œ$ there
is an infinite number of generations of ghosts; the SL(N|1) symmetry
might simplify their classification.

ÜACKNOWLEDGMENTS

We thank Gordon Chalmers, Vladimir Korepin, Olaf Lechtenfeld, Martin
Ro×cek, Cumrun Vafa, and Peter van Nieuwenhuizen for discussions.  N.B.
thanks the ITP at SUNY at Stony Brook for its hospitality where part of
this work was done. This work was supported by grant number
96/05524-0 of FAPESP.

\refs

£1 E.S. Fradkin and A.A. Tseytlin, \PL 106B (1981) 63;\\
	A. D'Adda and F. Lizzi, \PL 191B (1987) 85.

£2 M. Ademollo, L. Brink, A. D'Adda, R. D'Auria, E. Napolitano, S. Sciuto,
	E. Del Giudice, P. Di Vecchia, S. Ferrara, F. Gliozzi, R. Musto,
	R. Pettorino, and J.H. Schwarz, \NP 111 (1976) 77.

£3 H. Ooguri and C. Vafa, ÓMod. Phys. Lett.Õ ÉA5 (1990) 1389,
	\NP 361 (1991) 469, É367 (1991) 83.

£4 N. Marcus, \NP 387 (1992) 263.

£5 W. Siegel, \PR 69 (1992) 1493.

£6 M. Ademollo, L. Brink, A. D'Adda, R. D'Auria, E. Napolitano, S. Sciuto,
	E. Del Giudice, P. Di Vecchia, S. Ferrara, F. Gliozzi, R. Musto, and
	R. Pettorino, \NP 114 (1976) 297, \PL 62B (1976) 105.

£7 W. Siegel, \PRD 47 (1993) 2504.

£8 M. Ro×cek and G. Chalmers, private communications.

£9 N.J. Hitchin, A. Karlhede, U. Lindstr¬om, and M. Ro×cek,
	ÓCommun. Math. Phys.Õ É108 (1987) 535.

£10 W. Siegel, \PRD 46 (1992) R3235.

£11 G. Chalmers and W. Siegel, \PRD 54 (1996) 7628.

£12 N. Berkovits and C. Vafa, \NP 433 (1995) 123.

£13 R. Brooks (May, 1993), unpublished.

£14 A. Ashtekar, \PR 57 (1986) 2244, \PRD 36 (1987) 1587;\\
	T. Jacobson and L. Smolin, \PL 196B (1987) 39, 
	ÓClass. Quant. Grav.Õ É65 (1988) 583;\\
	J. Samuel, ÓPramanaÕ É28 (1987) L429.

£15 C.N. Yang, \PR 38 (1977) 1377.

£16 A.N. Leznov, ÓTheor. Math. Phys.Õ É73 (1988) 1233,\\
	A.N. Leznov and M.A. Mukhtarov, ÓJ. Math. Phys.Õ É28 (1987) 2574;\\
	A. Parkes, \PL 286B (1992) 265.

£17 J.F. Pleba«nski, ÓJ. Math. Phys.Õ É16 (1975) 2395.

£18 O. Lechtenfeld and W. Siegel, unpublished.

£19 N. Berkovits, \NP 450 (1995) 90.

£20 E. Witten, \NP 268 (1986) 253.

£21 H. Ooguri and C. Vafa, \NP 451 (1995) 121.

£22 J. Wess and B. Zumino, \PL 37B (1971) 95;\\
	S.P. Novikov, ÓSov. Math. Dok.Õ É24 (1981) 222;\\
	E. Witten, ÓCommun. Math. Phys.Õ É92 (1984) 455;\\
	S. Donaldson, ÓProc. Lond. Math. Soc.Õ É50 (1985) 1;\\
	V.P. Nair and J. Schiff, \PL 246B (1990) 423.

£23 S.J. Gates, Jr., M.T. Grisaru, M. Ro×cek, and W. Siegel, ÓSuperspaceÕ or 
	 ÓOne thousand and one lessons in supersymmetryÕ
	(Benjamin/Cummings, Reading, 1983) p. 173.

£24 E.P. Wigner, ÓAnnals Math.Õ É40 (1939) 149.

\bye